# Hyper-aging Dynamics of Nano-clay Suspension

A.Shahin and Yogesh M Joshi*

Department of Chemical Engineering, Indian Institute of Technology Kanpur,

Kanpur 208016. INDIA.

* E-Mail: joshi@iitk.ac.in

**Abstract**

Aqueous suspension of nanoclay Laponite undergoes structural evolution as a function of time, which enhances its elasticity and relaxation time. In this work we employ effective time approach to investigate long term relaxation dynamics by carrying out creep experiments. Typically we observe that the monotonic evolution of elastic modulus shifts to lower aging times while maxima in viscous modulus gets progressively broader for experiments carried out on a later date since preparation (idle time) of nanoclay suspension. Application of effective time theory produces superposition of all the creep curves irrespective of their initial state. The resulting dependence of relaxation time on aging time shows very strong hyper aging dynamics at small idle times, which progressively weakens to demonstrate linear dependence in the limit of very large idle times. Remarkably this behavior of nanoclay suspension is akin to that observed for polymeric glasses. Consideration of aging as a first order process suggests that continued hyper-aging dynamics causes cessation of aging. The dependence of relaxation time on aging time, therefore, must attenuate eventually producing linear or weaker dependence on time in order to approach progressively low energy state in the limit of very large times as observed experimentally. We also develop a simple scaling model based on a concept of aging of an energy well, which qualitatively captures various experimental observations very well leading to profound insight into the hyper-aging dynamics of nano-clay suspensions.

## I. Introduction

Clays are ubiquitous in nature and industry. Particularly 2:1 smectite clays find applications as rheology modifiers in health care, personal care, petroleum, paint, etc., industries due to their nanometric size, anisotropic (sheet-like) shape, and affordable cost. Common examples of such clay minerals are, Montmorrillonite, Bentonite, Laponite, etc.; which have layer thickness of around 1 nm with aspect ratio in the range 25 – 1000.[1, 2] These clay minerals are highly hydrophilic and swell in water. Typically, upon addition of clay in water, suspension slowly transforms itself from a flowing liquid to highly viscous soft solid (that supports its weight), suggesting presence of jamming.[3-8] Such transformation is also usually accompanied by evolution of its structure and various physical properties as a function of time.[9, 10] In this work we employ effective time framework to study jamming transition and relaxation dynamics of aqueous Laponite suspension using rheological tools. We also estimate relaxation time dependence on time elapsed since their preparation, which shows striking similarity with that of molecular glasses.

Synthetic hectorite clay, Laponite has a chemical formula: $(Na_{+0.7}[(Si_8Mg_{5.5}Li_{0.3})O_{20}(OH)_4]_{-0.7})$.[11] Laponite has a disc like shape with thickness 1 nm and diamter 25 - 30 nm and is farely monodispersed.[11, 12] Unit crystal of Laponite is composed of octahedral magnesia sheet sandwiched between two tetrahedral silica sheets. Isomorphic substitution of divalent magnesium by monovalent lithium creats paucity of positive charge within the layer, which is compensated by positive sodium ions present on the face of Laponite particle. On the edge of Laponite particles, where crystal structure is broken, presence of amphoteric groups, render the same positive charge at low pH while negative charge at high pH.[13] At pH 10 Laponite particles are believed to have positive edge charge.[14] In an aqueous medium, sodium ions dissociate giving the face of Laponite a permanant negative charge. In an aqueous

Laponite suspension competition between repulsion originating from negatively charged faces and attraction between negatively charged face and positively charged edge control the phase behavior.[15, 16] In the concentrated Laponite suspension (typically beyond 2 weight %), based on spectroscopic studies, some groups propose microstructure to be dominated by repulsion leading to a repulsive or Wigner glass.[17-19] Laponite particles do not physically touch each other in a Wigner glass and remain self suspended in the aqueous media due to repulsion. The other school claims, phase behavior of Laponite suspension to be dominated by attractive interactions among Laponite particles leading to a gel.[20-22] At intermediate Laponite concnetrations (1.1 to 2.4 wt %), one of the proposals suggests that the system can evolve as either a glass or a gel depending upon the state of the system owing to complex energy landscape of the same.[23] In addition to these claims, aqueous suspension of Laponite often demonstrates anisotropic phase,[24-26] which recently has been reported to have a pronounced presence near the air interface.[13]

Macroscopically, addition of Laponite in water leads to increase in viscosity as a function of elapsed time so that apparently flowing liquid gets converted to soft pasty material in a duration ranging from minutes to hours depending on the concentration of Laponite.[9, 27] Depending upon microstructure of the suspension, whether a repulsive glass or an attractive gel, such transformation can be represented as either a glass transition or gelation. Without going into complications of terminology we address this transformation by a generic term: jamming transition. Beyond the jamming transition, viscosity and elastic modulus of Laponite suspension continues to increase as a function of time, a phenomenon usually known as physical aging.[28] Rheologically such transformation is analyzed by studying evolution $G'$ and $G''$. For experiments performed on comparatively young sample, subsequent to shear melting, $G''$ is higher than $G'$ and the material exhibits a liquid like response. Eventually $G'$ crosses over $G''$,[15, 27] and the corresponding time marks the jamming transition.[29, 30] During the ageing process, relaxation time of the material also shows enhancement as a function of time. Typically, if dependence of relaxation time on age is weaker than linear, phenomenon is characterised as sub-aging,[31, 32] while if the same is stronger than linear the phenomenon is addressed as hyper-aging.[33, 34] Application of strong deformation field in a rheological experiment reverses the aging by causing "shear induced" melting of soft solid producing liquid. However, in case of Laponite suspension, the structural evolution is observed not to be completely reversible over very long durations of the time elapsed since the preparation of sample (timescales of days).[15] Typically shear melting of progressively older samples is observed to produce a liquid with greater viscosity.

Aqueous Laponite suspension, by virtue of its time dependent physical behavior, does not follow time – translational invariance[35, 36] Such behavior, therefore, does not let the material obey linear viscoelastic principles. Recently our group proposed a methodology based on an effective time theory and showed that linear viscoelastic principles can be successfully applied to such time dependent materials.[10, 35, 37] Application of effective time approach directly leads to estimation of dependence of relaxation time on waiting time. In this work we use effective time theory to study evolution of aqueous suspension of Laponite using rheological tools that gives new insights into the long term evolution of a soft glassy material.

**II. Material and Experimental Protocol:**

In this work we have used an aqueous suspension of 2.8 weight percent Laponite RD (Southern Clay products Inc.) having 0.5 weight % poly ethylene oxide (Loba Chemie, Mol. wt. 6000). Oven dried white powder of Laponite RD and PEO were added simultaneously to deionized water having pH 10 (maintained by addition of NaOH) and mixed using ultra turrex drive for a period of 45 minutes. The resulting suspension was stored in air tight polypropylene bottles at temperature 30°C. We denote the period of storage as an idle time ($t_i$) of the system usually denoted in days. In this work we carry out creep experiments using a stress controlled rheometer: AR 1000. On a given idle time the suspension was loaded in concentric cylinder geometry with an inner diameter 28 mm and a gap of 1 mm. To ensure a uniform initial state on a given idle time, the suspension was shear melted by applying oscillatory shear having stress magnitude 80 Pa at 0.1 Hz frequency for 20 min. Subsequently, in an aging step, evolution of elastic and viscous modulus of the suspension was monitored by applying small amplitude oscillatory shear with stress magnitude 1 Pa and frequency 0.1 Hz. Aging experiments were carried out for a predetermined time, known as aging time, following which creep experiments were carried out by applying constant stress having magnitude 5 Pa. In order to avoid drying, the free surface of the sample was covered with a thin layer of low viscosity silicon oil. All the experiments were carried out at 25°C.

It should be noted that addition of polyethylene oxide to aqueous Laponite suspensions can alter the ageing dynamics of the suspension and hence the ageing behaviour of Laponite-PEO suspensions could be different as compared to pure Laponite suspensions. Addition of low molecular weight PEO slows down the ageing dynamics of suspension due to steric hindrance,[38] whereas addition of high molecular weight PEO forms shake gels as the PEO chains have an ability to form bridges between Laponite platelets.[38, 39]

**III. Results and discussion:**

In figure 1 evolution of elastic and viscous modulus of shear melted Laponite suspension are plotted as a function of time, for experiments carried out on various days after preparation of suspension (idle time, $t_i$). It can be seen that the evolution of $G'$, which is monotonic and has self-similar curvature, shifts to smaller times for experiments carried on a later date since preparation (idle time). $G''$, on the other hand, demonstrates a maximum whose breadth becomes progressively broader at later idle times. We carry out such aging experiments on various idle times for predetermined aging times ($t_w$) and apply constant stress. The consequent creep curves obtained on a 21 day old Laponite suspension at various $t_w$ are plotted in the inset of figure 2. It can be seen that lesser compliance gets induced for experiments carried out at higher $t_w$. The aging time dependent creep flow behavior implies inapplicability of time-translational invariance (TTI). Therefore Boltzmann superposition principle is not applicable to the system. This principle is mathematically represented as:[40]

$$\gamma(t) = \int_{-\infty}^{t} J(t-t_w) \frac{d\sigma}{dt_w} dt_w , \qquad (1)$$

where $\gamma$ is strain in the material at present time $t$ under application of stress $\sigma$ applied at time $t_w$. $J$ is compliance, which solely depends on $t-t_w$. However, as shown in the inset of figure 2, for aqueous suspension of Laponite, $J$ shows an additional dependence on $t_w$ leading to: $J = J(t-t_w, t_w)$ invalidating Boltzmann superposition principle. This behavior is normally observed when relaxation time increases as a function of aging time. Under such circumstances it is customary to replace real time by an effective time obtained by rescaling the time dependent relaxation process with a constant relaxation time.[36] Effective time is defined as:[35, 36]

$$\xi(t) = \int_{0}^{t} \tau_0 dt' / \tau(t') , \qquad (2)$$

where $\tau$ is relaxation time, while $\tau_0$ is constant relaxation time associated with effective time. In effective time domain, since relaxation time remains constant, equation 1 can be applied to the present system by replacing $(t-t_w)$ by $(\xi(t)-\xi(t_w))$ in the expression for compliance: $J(t-t_w) = J(\xi(t)-\xi(t_w))$. However in order to estimate effective time, according to equation (2), it is necessary to know the dependence of relaxation time on aging time: $\tau(t')$. It is usually observed that aqueous suspension of Laponite, when it has soft solid like consistency, follows power law dependence of relaxation time on aging time given by: $\tau = A\tau_m^{1-\mu}t'^{\mu}$,[28, 35, 36] where $\tau_m$ is microscopic timescale associated with the suspension (discussed below), $A$ is a constant pre-factor while $\mu$ is logarithmic rate of change of relaxation time with respect to aging time. When Laponite suspension in the liquid state $(G' < G'')$, it is observed to follow exponential dependence on time given by: $\tau(t') = \tau_m \exp(\alpha t')$.[35] [Aging dynamics of freshly prepared Laponite suspensions studied using dynamic light scattering experiments also show an initial exponential growth followed by a linear growth of relaxation time.[41-43] However we believe that this behavior is related to cage formation dynamics[5] and is different compared to that discussed in the present work wherein experiments are carried out on atleast 13 day old shear melted Laponite suspension]. In figure 2 we have plotted creep curves associated with 21 day old Laponite suspension over the duration of aging times in the range 1800 to 14400 s. As shown in figure 1, over this duration, elastic modulus is greater than viscous modulus suggesting sample to be in a solid like state. For power law dependence, effective time elapsed since application of deformation field according to equation (2) is given by:

$$\xi(t) - \xi(t_w) = \tau_0 \tau_m^{\mu-1}\left[t^{1-\mu} - t_w^{1-\mu}\right]/[A(1-\mu)]. \tag{3}$$

If we assume $\tau_0 = \tau(t_{wm})$, where $t_{wm}$ is maximum aging time (14400 s) employed in the experiments, we get:

$$\xi(t) - \xi(t_w) = t_{wm}^{\mu}\left[t^{1-\mu} - t_w^{1-\mu}\right]/(1-\mu). \tag{4}$$

It can be seen that creep compliance demonstrates an excellent superposition for a certain unique value of $\mu$, when plotted against $\left[t^{1-\mu} - t_w^{1-\mu}\right]/(1-\mu)$ as shown in figure 2. The unique value of $\mu$ necessary to obtain superposition, suggests dependence of relaxation time on aging time given by: $\mu = d\ln\tau/d\ln t_w$. Figure 2, therefore, successfully validates Boltzmann superposition principle in the effective time domain. Importantly, for soft glassy materials including nanoclay suspensions, Boltzmann superposition principle in an effective time domain not just gets validated for experiments performed at different aging times but is also observed to get validated for different stress histories and different temperature histories as discussed in the literature.[10, 35, 37] It should be noted that procedure for estimation of $\mu$ using effective time theory is different from the conventional procedure.[28, 44-46] In the conventional technique, which is due to Struik,[28] creep data is considered over only 10 % of aging time in order to avoid effect of aging during the course of creep flow. Value of $\mu$ is obtained from a superposition by horizontally shifting the creep data. The effective time procedure, on the other hand, allows use of creep data in its entirety to produce a superposition for a unique value of $\mu$.

We carried out the creep experiments as shown in the inset of figure 2 at various idle times in the range 13 days to 180 days. Similar to that shown in figure 2, creep curves associated with various idle times also demonstrate superposition when plotted against ($\left[t^{1-\mu} - t_w^{1-\mu}\right]/(1-\mu)$), which we have described in figure 3. Unlike all the explored idle times, experiments carried out on $t_i = 13$ days demonstrate exponential dependence of relaxation time on aging time: $\tau(t') = \tau_m \exp(\alpha t')$. This is because the sample on day 13 is observed to be in the liquid state ($G' < G''$) over the most of explored aging times. According to equation (2), effective time elapsed since application of deformation for exponential dependence is given by: $\xi(t) - \xi(t_w) = (\tau_0/\tau_m)\left[\exp(-\alpha t_w) - \exp(-\alpha t)\right]/\alpha$. If we assume $\tau_0 = \tau(t_{wm})$,

$$\xi(t) - \xi(t_w) = \left(\exp(\alpha t_{wm})\right)\left[\exp(-\alpha t_w) - \exp(-\alpha t)\right]/\alpha. \tag{5}$$

We have plotted normalized compliance against $\left[\exp(-\alpha t_w)-\exp(-\alpha t)\right]/\alpha$ in figure 3(a) for day 13 creep data. For all the other explored idle times (21 days and beyond), samples show power law dependence. The creep curves on various idle times demonstrate superposition as shown in figure 3(a) for unique values of $\mu$ that depend only on $t_w$. In figure 4, we plot $\mu$ as a function of idle time. It can be seen that in the initial idle time region, $\mu$ is significantly greater than unity suggesting hyper aging dynamics. With increase in idle time, $\mu$ decreases rapidly and in the limit of high idle times reaches a plateau around unity suggesting a linear dependence of relaxation time on aging time.

All the superpositions plotted in figure 3(a) have self-similar curvature. The superpositions can therefore be shifted horizontally to obtain a comprehensive superposition as shown in figure 3(b). In figure 3(a), individual superposition at each idle time is plotted as a function of $\left[\xi(t)-\xi(t_w)\right]t_{wm}^{-\mu}$ (For day 13 data, superposition is plotted against $\left[\xi(t)-\xi(t_w)\right]\exp(-\alpha t_{wm})$). The constant relaxation time $(\tau_0)$ associated with each superposition is the relaxation time associated with maximum aging time at that idle time: $\tau_0 = \tau(t_{wm}, t_i)$. Therefore, if the comprehensive superposition does exist, then the horizontal shift factor is equal to $\tau(t_{wm}, t_i)/\tau(t_{wm}, t_{iR})$, where $t_{iR}$ is reference idle time. In figure 5, we plot $\tau(t_{wm}, t_i)/\tau(t_{wm}, t_{iR})$ as a function of idle time ($t_{iR}$=45 days). We also plot evolutions of elastic and viscous moduli associated with few idle times as insets on the same plot. Overall, the relaxation time increases with increase in idle time, however the maximum change is observed between points associated with 13 day and 21 day data. On the day 13, sample was in liquid state $(G' < G'')$ at those aging times when creep experiments were carried out. On the other hand, on day 21 sample was in solid state $(G' > G'')$ when creep experiments were carried out. The jump in relaxation time over three decades between day 13 and 21, therefore, suggests liquid – solid jamming transition in aqueous suspension of Laponite. At greater idle times, when system is always in the solid state (when creep experiments were carried out), relaxation time shows comparatively weaker enhancement.

It is important to note that the hyper-aging behavior observed in the present work should not be confused with hyper-diffusion reported for aqueous Laponite suspension. For example, on one hand Bandyopadhyay and coworkers[34] observed dependence of relaxation time on aging time to be stronger than linear ($\mu$ =1.8), on the other hand they reported faster than exponential relaxation at explored aging times. The former phenomenon is represented as hyper-aging, while the latter is termed as hyper-diffusion. In the literature, hyper-diffusive behavior has been observed for Laponite suspension even though hyper-aging is absent.[42, 43] We therefore believe that hyper-aging and hyper-diffusion phenomena are unrelated with latter caused by ballistic movement associated with elastic deformation originating from heterogeneous local stress.[47] In case of glassy materials scaling arguments suggest relaxation time scales as aging time as this is the only time scale available to the system. The very fact that glassy material shows stronger than linear relationship suggest that processes other than simple physical aging with different timescales are influencing the dynamics. We therefore believe that irreversible dynamics shown by aqueous Laponite suspension might be related to observed hyper-aging behavior. However, we feel that further work is necessary in order understand physical origin of the same.

Time dependency associated with aqueous suspension of Laponite is related to thermodynamically out – of – equilibrium character of the same. Such time dependent behavior, typically addressed as physical aging, is also observed in many out - of - equilibrium systems such as soft glassy materials,[35, 48] spin glasses,[31, 49] polymeric glasses,[28, 50, 51] etc. A simple observation that addition of around 1.1 volume % of Laponite discs (≈ 2.8weight %) enhances viscosity of water over 5 orders of magnitude suggests that translational diffusivity of the nanometric particles is significantly hindered. Owing to such reduced mobility, only limited part of the phase space is accessible to the system. Nonetheless, highly constrained motion of the jammed Laponite particles causes slow but steady rearrangement of the structure taking the system to progressively lower energy state. This concept can be understood by representing energetic interaction of individual Laponite particles with its neighbors by a potential energy well. [Please refer to Figure S1 of Supplementary information]. In the out of equilibrium state individual particle is only occasionally able to jump out of energy well due to limited thermal energy associated with the same. In addition, the particle is free to undergo microscopic motion while remaining surrounded by the same neighbors.[34] Overall the structural rearrangement prefers those configurations that decrease energy of the particle as a function time. The elastic modulus of the system, in such case can then be obtained as suggested by Jones.[52] If we represent inter-particle interactions by Hookean (quadratic) springs with spring constant $k$, and if $b$ is the average inter-particle distance, then elastic modulus is given by: $k/b$. The spring constant $k$ is equal to second derivative of the interaction potential computed at the point of minimum energy in the well.[52] By using simple scaling arguments, Jones[52] suggested that $k \sim E/b^2$. If we assume all the particles in the system are trapped in wells with the same energy barrier, elastic modulus can be represented by:

$$G' = \beta E/b^3 , \qquad (6)$$

where $\beta$ is a constant of proportionality. This concept can also be understood from schematic of energy well shown in figure S1 of Supplementary information. The displacement of the particle by distance $r$ enhances energy of the particle by $\tfrac{1}{2}kr^2$. In the aging process well depth goes on increasing as a function of time, however average inter-particle distance $b$ can be expected to remain unaffected. Consequently, in order to have same deformation, greater energy is necessary ($\tfrac{1}{2}kb^2 \approx E$), which again leads to: $G' = \beta E/b^3$. Therefore, a simple picture of aging, wherein energy well depth increases as a function of time predicts enhancement of elastic modulus as described in figure S1 of Supplementary information.

Knowledge of depth of energy well in which particles are trapped also leads to characteristic relaxation time through Arrhenius relationship.[36] Interestingly relaxation time is observed to show power law dependence on aging time in the jammed state.[28] Equivalency of both the equations leads to:

$$\tau = \tau_m \exp(E/k_B T) = A\tau_m (t_w/\tau_m)^{\mu}. \tag{7}$$

Equation 7 models the relaxation time as a cage diffusion time. In soft glassy rheology framework cage diffusion is considered as an activated process.[53] Fielding *et al.*[36] describe the time scale $\tau_m$ as microscopic attempt time for the activation process. This timescales sets the rate at which material ages. Combining equations (6) and (7) leads to an expression for elastic modulus given by:

$$G' = \beta \frac{E}{b^3} = \frac{\beta k_B T}{b^3} \ln A + \mu \frac{\beta k_B T}{b^3} \ln(t_w/\tau_m) \tag{8}$$

Assuming Maxwell type behavior, viscous modulus can be related to elastic modulus by: $G'' = G'/\omega\tau$. Using equations 7 and 8, viscous modulus can be written as:

$$\ln G'' = -\ln(\omega A\tau_m) + \ln\left[\frac{\beta k_B T}{b^3}\ln A + \mu\frac{\beta k_B T}{b^3}\ln(t_w/\tau_m)\right] - \mu\ln(t_w/\tau_m). \tag{9}$$

Interestingly equation 9 qualitatively explains decrease in viscous modulus as a function of aging time at high times (last term) as shown in figure 1. It should be noted that, figure 1 shows that decrease in $G''$ is stronger for experiments carried out at smaller idle times. Equation 9 suggests that, on double logarithmic scale $G''$ decreases with slope $\mu$. Remarkably, decrease in $\mu$ at greater idle times as shown in figure 4 nicely compliments weakening slope of $G''$ at greater idle times as observed in figure 1.

Overall figure 5 discusses how the relaxation dynamics changes from very rapid exponential enhancement on day 13 to strong power law dependence over days 21 to 56 (hyper aging) to linear dependence in the limit of very long idle times. According to our simple model description, equation (8) suggests that energy well depth increases with respect to time as: $E \propto \mu k_B T \ln(t_w/\tau_m)$. Owing to greater value of $\mu$, enhancement of energy well depth of shear rejuvenated Laponite suspension is very rapid over the initial duration. In the limit of very large idle times, however, the rate of deepening of energy well becomes steady leading to linear relation on aging time. Usually for glassy materials linear dependence of relaxation time on aging time is considered to be a signature characteristic of the same. Struik,[28] based on overwhelming data on glassy polymers and also theoretical arguments, argued that factor $\mu$ should tend to unity in the limit of long times. Remarkably our experiments on Laponite suspension also suggest the same phenomenon.

The qualitative discussion explaining enhancement of elastic modulus, broadness of maxima of viscous modulus and behavior of $\mu$ is based on assumption that all the particles are trapped in energy wells having depth $E$. However, in reality there exists a distribution of well depths in which particles are arrested ($E_i$). Such distribution also leads to distribution in relaxation times through the dependence: $\tau_i = \tau_m \exp(E_i/k_B T)$. Owing to the aging dynamics, the particles lower their energy with time which causes slowing down of the relaxation dynamics. In figure 2, creep curves obtained at various aging times are shown to superpose when plotted in effective time domain. However, such superposition is possible only if all the relaxation modes evolve with the same dependence on aging time ($\mu$ for all the modes remains constant).[37] Consequently all the modes can be seen to be increase with aging time keeping shape of the spectrum constant. (Please refer to figure S2 of Supplementary information). In figure 3, time – aging time superpositions obtained on various idle times are shown to have self-similar curvatures which lead to a comprehensive superposition. In order to observe such superposition, it is again necessary that shape of relaxation time distribution should remain unaffected as a function of idle time. The typical procedure, which is followed during the experiments, involves shear melting (mechanical quench) to be carried out on the samples. Shear melting induces strong deformation field in the material fluidizing the jammed entities thereby enhancing their energy (reducing energy well depth). Subsequent to shear melting, particles occupy the energy wells with distribution of well depths. However, as shown in figures 1, shear melting carried out on greater idle times does not rejuvenate the sample to the same initial state. This may be due to certain structure formation which is too strong to be destroyed by the induced shear (energy wells are too deep compared to strain induced energy). Therefore, the relaxation time distribution subsequent to shear melting, on progressively higher idle times, shifts to deeper well depths preserving its shape.

Under constant temperature and constant pressure conditions, which is a case with the present experiments, evolution of structure takes place in order to lower the Gibbs free energy of the system[54] (Since suspension of Laponite is in aqueous media and ergodicity breaking occurs over particle length-scale, volume of suspension does not change while it ages. Consequently, the aging process in Laponite suspension is also a constant volume – constant temperature process. Therefore, lowering of Helmholtz free energy also characterizes the aging process[54]). Let's assume that at any idle time specific Gibbs free energy of the system immediately after the shear melting is stopped is $g_0$. On the other hand, let the lowest possible Gibbs free energy of the system (corresponding to the equilibrium state) be $g_\infty$. Therefore, during an aging process Gibbs free energy of a system ($g$) decreases from $g_0$ towards $g_\infty$ as a function of time. If evolution of Gibbs free energy is assumed to be a first order process, we have: [28, 55]

$$\frac{dg}{dt} = -\frac{g - g_\infty}{\tau(t)}. \tag{10}$$

Difference ($g - g_\infty$) can be represented as excess Gibbs free energy. The rate of change of relaxation time can then be related to the dependence of relaxation time on excess Gibbs free energy as:

$$\frac{d\tau(t)}{dt} = -\frac{d \ln \tau(t)}{d \ln(g - g_\infty)} = f[g(t)] \tag{11}$$

The logarithmic dependence of relaxation time on aging time can then be represented by:[28]

$$\mu = \frac{d \ln \tau}{d \ln t} = \frac{t\, f[g(t)]}{\int_0^t f[g(t')]dt' + \tau_b}, \tag{12}$$

where $\tau_b$ is relaxation time at the beginning of aging ($t = 0$ and $g = g_0$). Equation (12) can be easily rearranged to give:

$$\frac{1}{\mu} = \int_0^1 \varphi\, d\zeta + \frac{\tau_b}{t\, f[g(t)]}, \tag{13}$$

where $\varphi(t',t) = f[g(t')]/f[g(t)]$ and $\zeta = t'/t$. Therefore, as $\zeta$ approaches unity ($t' \to t$), $\varphi(t',t)$ also approaches unity ($\varphi \to 1$). The second term on the right hand side of equation 13 is expected to become negligible in the limit of large aging times. Therefore, in order to observe hyper-aging ($\mu > 1$), $\varphi$ must be an increasing function of $\zeta$ ($d\varphi/d\zeta > 1$). On the other hand, in order to observe linear dependence of relaxation time on aging time ($\mu \approx 1$), $\varphi$ must remain constant over greater duration of $\zeta$ ($d\varphi/d\zeta \approx 0$). As shown in figure 5, we observe relaxation time to follow power law dependence on aging time given by: $\tau = A\tau_m^{1-\mu}t^\mu$, which leads to:

$$\varphi = \zeta^{\mu-1}. \tag{14}$$

In figure 6 (a) we have plotted equation (14) describing behavior of $\varphi$ with respect to $\zeta$ for various values of $\mu$. In addition the dependence of relaxation time on excess Gibbs free energy for $\tau = A\tau_m^{1-\mu}t^\mu$ can be obtained from equation (11) as:

$$\begin{aligned} \frac{\tau_b^v}{\tau^v} &= 1 + (\mu-1)A^{1/\mu}\left(\frac{\tau_b}{\tau_m}\right)^v \ln\left(\frac{g-g_\infty}{g_0-g_\infty}\right) \quad \cdots \quad \mu > 1 \\ \frac{\tau}{\tau_b} &= \left(\frac{g-g_\infty}{g_0-g_\infty}\right)^{-A} \quad \cdots \quad \mu = 1 \end{aligned}, \tag{15}$$

where $v = (\mu-1)/\mu$. For values of $\mu > 1$, relaxation time diverges at $\ln\left[(g-g_\infty)/(g_0-g_\infty)\right] = \left[A^{-1/\mu}(\tau_m/\tau_b)^v/(1-\mu)\right]$ leading to cessation of aging. In figure 6(b) we have plotted equation (15) for various values $\mu$ by considering a special case of $A=1$ and $\tau_b = \tau_m$. Equation (15) and its description in figure 6(b) clearly suggests that the system can never approach an equilibrium state ($g = g_\infty$), with $\mu > 1$ (Interestingly $\mu >> 1$ limit can be considered as a granular limit, as microscopic rearrangement is frozen in granular media and the system remains in a disordered higher energy state indefinitely). On the other hand, for $\mu \leq 1$, aging continues and equilibrium state is approached in the limit of very large time ($t \to \infty$). Our observation, described in figure 4 suggests that dependence of relaxation time on excess Gibbs free energy changes from strongly diverging dependence ($\mu > 1$) at smaller idle times to a linear dependence at higher idle times. Therefore by slowly transforming relaxation dynamics from a hyper aging regime to linear regime, continuing aging dynamics is facilitated in aqueous Laponite suspension.

**IV. Conclusion:**

In this work we study long term relaxation dynamics of aqueous suspension of nano-clay Laponite through effective time theory. Oscillatory and creep experiments were performed on shear rejuvenated suspension samples up to 180 days after preparation of the same. Typically elastic modulus of suspension shows enhancement as a function of time. Viscous modulus, which increases in the beginning of the aging process, shows a decrease after crossing over elastic modulus. Experiments performed on later dates since preparation of suspension (idle time), cause evolution of $G'$ to shift to lower time scale while broadening the maxima for $G''$. Owing to enhanced relaxation time of the suspension, creep experiments induce lesser strain in the sample for experiments carried out at greater aging times thereby invalidating Boltzmann superposition principle. In order to apply this principle, we transform the real time to effective time by rescaling the time dependent relaxation processes by a constant relaxation time. Since relaxation time is constant in the effective time domain, Boltzmann superposition principle produces superposition of all the creep data. Superposition also leads to a relationship between relaxation time and aging time, which typically has power law type dependence in the glassy domain ($G'>G''$). We observe that the power law exponent $\mu$, which suggests logarithmic dependence of relaxation time on aging time, is significantly greater than unity (hyper-aging behavior) at small idle times. Value of $\mu$ sharply decreases with idle time and in the limit of large idle times, a linear dependence of relaxation time on aging time is observed ($\mu \approx 1$). Consideration of physical aging as a first order process suggests that for hyper aging dynamics ($\mu>1$), relaxation time diverges terminating the process of aging over finite time scales. Therefore, In order to have aging continue in the system so as to eventually get closer to equilibrium state, $\mu$ must approach unity. We indeed observe $\mu$ to approach unity at very large idle times. Interestingly, the observation of linear dependence in the limit of very long times is akin to that observed for glassy polymeric materials.

Even though the dependence of relaxation time on aging time changes for experiments carried out at greater idle time, superpositions obtained on various idle times show self-similar curvature. We obtain a comprehensive superposition by horizontally shifting individual superpositions obtained on various idle times. Existence of such comprehensive superposition suggests that the shape of relaxation time distribution remains unaffected during aging and rejuvenation process, though average value of relaxation time may undergo a change. We also propose a simple model that qualitatively explains various rheological observations very well.

**Acknowledgement:** Financial support from Department of Science Technology, Government of India through IRHPA scheme is greatly acknowledged.

**References:**

1.   Meunier, A., *Clays*. Springer: Berlin, 2005.
2.   Van Olphen, H., *An Introduction to Clay Colloid Chemistry*. Wiley: New York, 1977.
3.   Ruzicka, B.; Zaccarelli, E.; Zulian, L.; Angelini, R.; Sztucki, M.; Moussaïd, A.; Narayanan, T.; Sciortino, F., Observation of empty liquids and equilibrium gels in a colloidal clay. *Nat Mater* **2011,** *10*, 56-60.
4.   Ruzicka, B.; Zulian, L.; Zaccarelli, E.; Angelini, R.; Sztucki, M.; Moussaïd, A.; Ruocco, G., Competing Interactions in Arrested States of Colloidal Clays. *Phys. Rev. Lett.* **2010,** *104*, 085701.
5.   Joshi, Y. M., Model for cage formation in colloidal suspension of laponite. *J. Chem. Phys.* **2007,** *127*, 081102.
6.   Shalkevich, A.; Stradner, A.; Bhat, S. K.; Muller, F.; Schurtenberger, P., Cluster, Glass, and Gel Formation and Viscoelastic Phase Separation in Aqueous Clay Suspensions. *Langmuir* **2007,** *23*, 3570-3580.

7. Cummins, H. Z., Liquid, glass, gel: The phases of colloidal Laponite. *J. Non-Cryst. Solids* **2007,** *353*, 3891-3905.
8. Bonn, D.; Kellay, H.; Tanaka, H.; Wegdam, G.; Meunier, J., Laponite: What is the difference between a gel and a glass. *Langmuir* **1999,** *15*, 7534-7536.
9. Joshi, Y. M.; Reddy, G. R. K.; Kulkarni, A. L.; Kumar, N.; Chhabra, R. P., Rheological Behavior of Aqueous Suspensions of Laponite: New Insights into the Ageing Phenomena. *Proc. Roy. Soc. A* **2008,** *464*, 469-489.
10. Gupta, R.; Baldewa, B.; Joshi, Y. M., Time Temperature Superposition in Soft Glassy Materials. *Soft Matter, DOI:10.1039/C2SM07071E.* **2012**.
11. http://www.laponite.com.
12. Kroon, M.; Vos, W. L.; Wegdam, G. H., Structure and formation of a gel of colloidal disks. *Phys. Rev. E* **1998,** *57*, 1962-1970.
13. Shahin, A.; Joshi, Y. M.; Ramakrishna, S. A., Interface-Induced Anisotropy and the Nematic Glass/Gel State in Jammed Aqueous Laponite Suspensions. *Langmuir* **2011,** *27*, 14045–14052.
14. Tawari, S. L.; Koch, D. L.; Cohen, C., Electrical double-layer effects on the Brownian diffusivity and aggregation rate of Laponite clay particles. *J.Colloid Interface Sci.* **2001,** *240*, 54-66.
15. Shahin, A.; Joshi, Y. M., Irreversible Aging Dynamics and Generic Phase Behavior of Aqueous Suspensions of Laponite. *Langmuir* **2010,** *26*, 4219–4225.
16. Ruzicka, B.; Zaccarelli, E., A fresh look at Laponite phase diagram. *Soft Matter* **2011,** *7*, 1268-1286.
17. Jabbari-Farouji, S.; Wegdam, G. H.; Bonn, D., Gels and Glasses in a Single System: Evidence for an Intricate Free-Energy Landscape of Glassy Materials. *Phys. Rev. Lett.* **2007,** *99*, 065701-4.
18. Ruzicka, B.; Zulian, L.; Angelini, R.; Sztucki, M.; Moussaid, A.; Ruocco, G., Arrested state of clay-water suspensions: Gel or glass? *Phys. Rev. E* **2008,** *77*, 020402-4.
19. Mourchid, A.; Delville, A.; Lambard, J.; Lecolier, E.; Levitz, P., Phase diagram of colloidal dispersions of anisotropic charged particles: Equilibrium properties, structure, and rheology of laponite suspensions. *Langmuir* **1995,** *11*, 1942-1950.
20. Mongondry, P.; Tassin, J. F.; Nicolai, T., Revised state diagram of Laponite dispersions. *J.Colloid Interface Sci.* **2005,** *283*, 397-405.
21. Sun, K.; Kumar, R.; Falvey, D. E.; Raghavan, S. R., Photogelling Colloidal Dispersions Based on Light-Activated Assembly of Nanoparticles. *J. Am. Chem. Soc.* **2009,** *131*, 7135-7141.
22. Dijkstra, M.; Hansen, J.-P.; Madden, P. A., Statistical model for the structure and gelation of smectite clay suspensions. *Phys. Rev. E* **1997,** *55*, 3044-3053.
23. Jabbari-Farouji, S.; Tanaka, H.; Wegdam, G. H.; Bonn, D., Multiple nonergodic disordered states in Laponite suspensions: A phase diagram. *Phys. Rev. E* **2008,** *78*, 061405-10.

24. Gabriel, J.-C. P.; Sanchez, C.; Davidson, P., Observation of nematic liquid-crystal textures in aqueous gels of smectite clays. *J. Phys. Chem.* **1996,** *100*, 11139-11143.
25. Lemaire, B. J.; Panine, P.; Gabriel, J. C. P.; Davidson, P., The measurement by SAXS of the nematic order parameter of laponite gels. *Europhys. Lett.* **2002,** *59*, 55-61.
26. Michot, L. J.; Bihannic, I.; Maddi, S.; Funari, S. S.; Baravian, C.; Levitz, P.; Davidson, P., Liquid-crystalline aqueous clay suspensions. *Pro. Nat. Acad. Sci.* **2006,** *103*, 16101-16104.
27. Willenbacher, N., Unusual thixotropic properties of aqueous dispersions of Laponite RD. In *J.Colloid Interface Sci.*, 1996; Vol. 182, pp 501-510.
28. Struik, L. C. E., *Physical Aging in Amorphous Polymers and Other Materials*. Elsevier: Houston, 1978.
29. Ovarlez, G.; Coussot, P., Physical age of soft-jammed systems. *Phys. Rev. E* **2007,** *76*, 011406.
30. Negi, A. S.; Osuji, C. O., Time-resolved viscoelastic properties during structural arrest and aging of a colloidal glass. *Physical Review E 82*, 031404.
31. Sibani, P.; Kenning, G. G., Origin of end-of-aging and subaging scaling behavior in glassy dynamics. *Physical Review E* **2010,** *81*, 011108.
32. Rinn, B.; Maass, P.; Bouchaud, J.-P., Hopping in the glass configuration space: Subaging and generalized scaling laws. *Physical Review B* **2001,** *64*, 104417.
33. Bissig, H.; Romer, S.; Cipelletti, L.; Trappe, V.; Schurtenberger, P., Intermittent dynamics and hyper-aging in dense colloidal gels. *PhysChemComm* **2003,** *6*, 21-23.
34. Bandyopadhyay, R.; Liang, D.; Yardimci, H.; Sessoms, D. A.; Borthwick, M. A.; Mochrie, S. G. J.; Harden, J. L.; Leheny, R. L., Evolution of particle-scale dynamics in an aging clay suspension. *Phys. Rev. Lett.* **2004,** *93*.
35. Shahin, A.; Joshi, Y. M., Prediction of long and short time rheological behavior in soft glassy materials. *Phys. Rev. Lett.* **2011,** *106*, 038302.
36. Fielding, S. M.; Sollich, P.; Cates, M. E., Aging and rheology in soft materials. *J. Rheol.* **2000,** *44*, 323-369.
37. Baldewa, B.; Joshi, Y. M., Delayed Yielding in Creep, Time - Stress Superposition and Effective Time Theory for a soft Glass. *Soft Matter* **2012,** *8*, 789-796.
38. Mongondry, P.; Nicolai, T.; Tassin, J.-F., Influence of pyrophosphate or polyethylene oxide on the aggregation and gelation of aqueous laponite dispersions. *Journal of Colloid and Interface Science* **2004,** *275*, 191-196.
39. Zebrowski, J.; Prasad, V.; Zhang, W.; Walker, L. M.; Weitz, D. A., Shake-gels: shear-induced gelation of laponite-PEO mixtures. *Colloids and Surfaces A: Physicochemical and Engineering Aspects* **2003,** *213*, 189-197.
40. Bird, R. B.; Armstrong, R. C.; Hassager, O., *Dynamics of Polymeric Liquids, Fluid Mechanics*. Wiley-Interscience: New York, 1987.

41. Schosseler, F.; Kaloun, S.; Skouri, M.; Munch, J. P., Diagram of the aging dynamics in laponite suspensions at low ionic strength. *Phys. Rev. E* **2006,** *73*, 021401.
42. Kaloun, S.; Skouri, R.; Skouri, M.; Munch, J. P.; Schosseler, F., Successive exponential and full aging regimes evidenced by tracer diffusion in a colloidal glass. *Phys. Rev. E* **2005,** *72*, 011403.
43. Bellour, M.; Knaebel, A.; Harden, J. L.; Lequeux, F.; Munch, J.-P., Aging processes and scale dependence in soft glassy colloidal suspensions. *Phys. Rev. E* **2003,** *67*, 031405.
44. Awasthi, V.; Joshi, Y. M., Effect of temperature on aging and time–temperature superposition in nonergodic laponite suspensions. *Soft Matter* **2009,** *5*, 4991–4996.
45. Joshi, Y. M.; Reddy, G. R. K., Aging in a colloidal glass in creep flow: Time-stress superposition. *Phys. Rev. E* **2008,** *77*, 021501-4.
46. Cloitre, M.; Borrega, R.; Leibler, L., Rheological aging and rejuvenation in microgel pastes. *Phys. Rev. Lett.* **2000,** *85*, 4819-4822.
47. Cipelletti, L.; Ramos, L.; Manley, S.; Pitard, E.; Weitz, D. A.; Pashkovski, E. E.; Johansson, M., Universal non-diffusive slow dynamics in aging soft matter. *Faraday Discussions* **2003,** *123*, 237-251.
48. Agarwal, P.; Qi, H.; Archer, L. A., The Ages in a Self-Suspended Nanoparticle Liquid. *Nano Letters* **2009,** *10*, 111-115.
49. Sibani, P.; Hoffmann, K. H., Hierarchical-models for aging and relaxation of spin-glasses. *Physical Review Letters* **1989,** *63*, 2853-2856.
50. Hodge, I. M., Physical aging in polymer glasses. *Science* **1995,** *267*, 1945-1947.
51. Guo, Y.; Zhang, C.; Lai, C.; Priestley, R. D.; D'Acunzi, M.; Fytas, G., Structural Relaxation of Polymer Nanospheres under Soft and Hard Confinement: Isobaric versus Isochoric Conditions. *ACS Nano* **2011,** *5*, 5365-5373.
52. Jones, R. A. L., *Soft Condensed Matter*. Oxford University Press: Oxford, 2002.
53. Sollich, P.; Lequeux, F.; Hebraud, P.; Cates, M. E., Rheology of soft glassy materials. *Phys. Rev. Lett.* **1997,** *78*, 2020-2023.
54. Callen, H. B., *Thermodynamics and an introduction to thermostatistics*. John Wiley & Sons: New York, 1985; p 13.
55. Shaw, M. T.; MacKnight, W. J., *Introduction to Polymer Viscoelasticity*. 3 edition ed.; Wiley: New York, 2005.

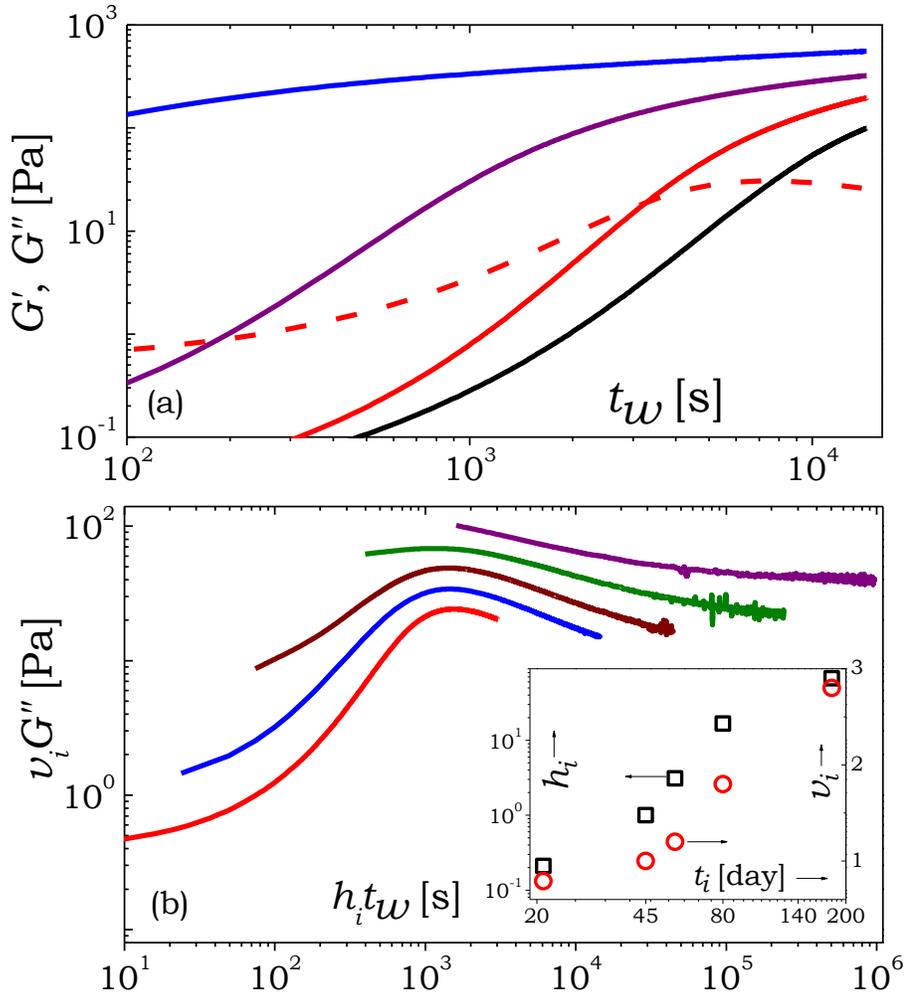

**Figure 1.** Evolution of elastic modulus and viscous modulus (a) Elastic modulus (full line) is plotted as a function of aging time for experiments carried out on various days elapsed after preparation of Laponite suspension (idle time, $t_i$). From right to left: black line $t_i$ =13 day, red line: 21 day, purple line: 45 day, blue line: 80 day. We have also plotted $G''$ for a 21 days data as a dashed red line. (b) Viscous modulus is plotted as a function of aging time (From top to bottom: 180, 80, 56, 45, 21 day). The curves are shifted vertically and horizontally for clarity. It can be seen that with increase in idle time maxima in $G''$ becomes broad. Inset shows the horizontal shift factors (black squares) and vertical shift factors (red circles) used to demarcate the curvature of $G''$.

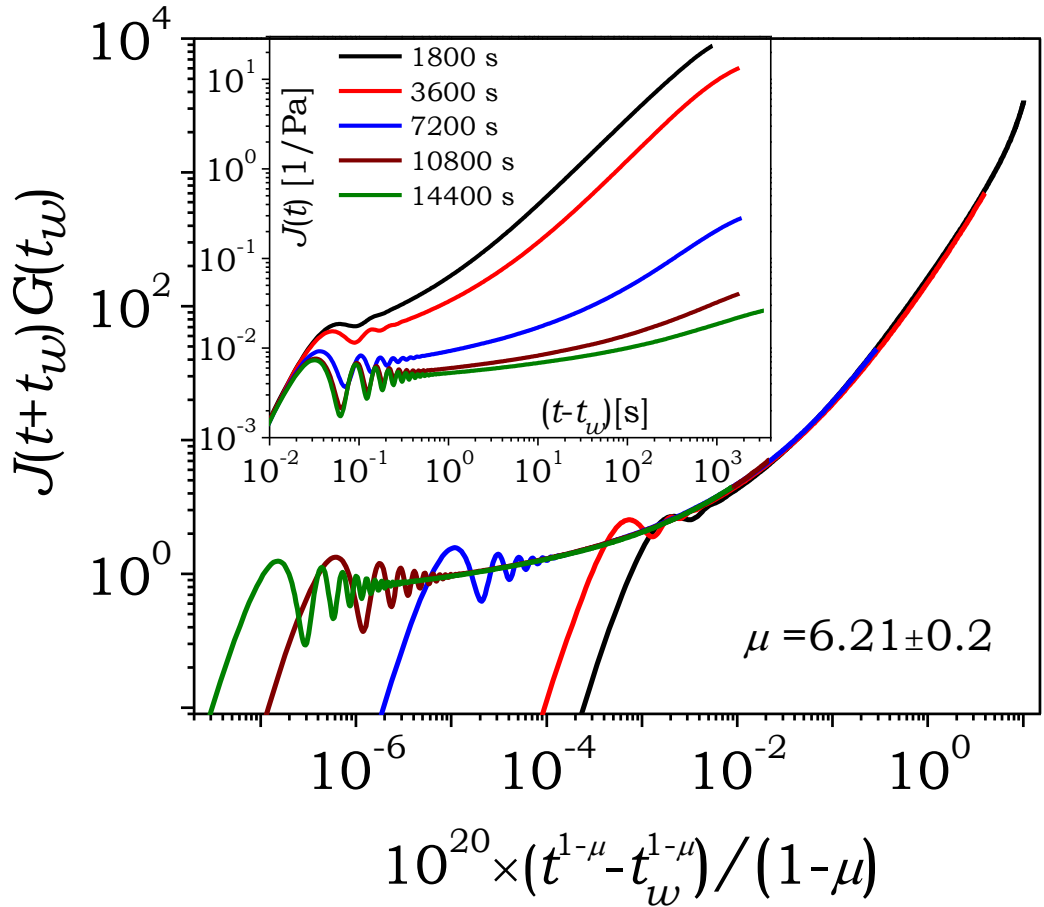

**Figure 2.** Time-aging time superposition for creep curves generated on $t_i$ = 21 day. The inset shows creep curves obtained at various waiting times (from top to bottom: 1800, 3600, 7200, 10800, and 14400 s).

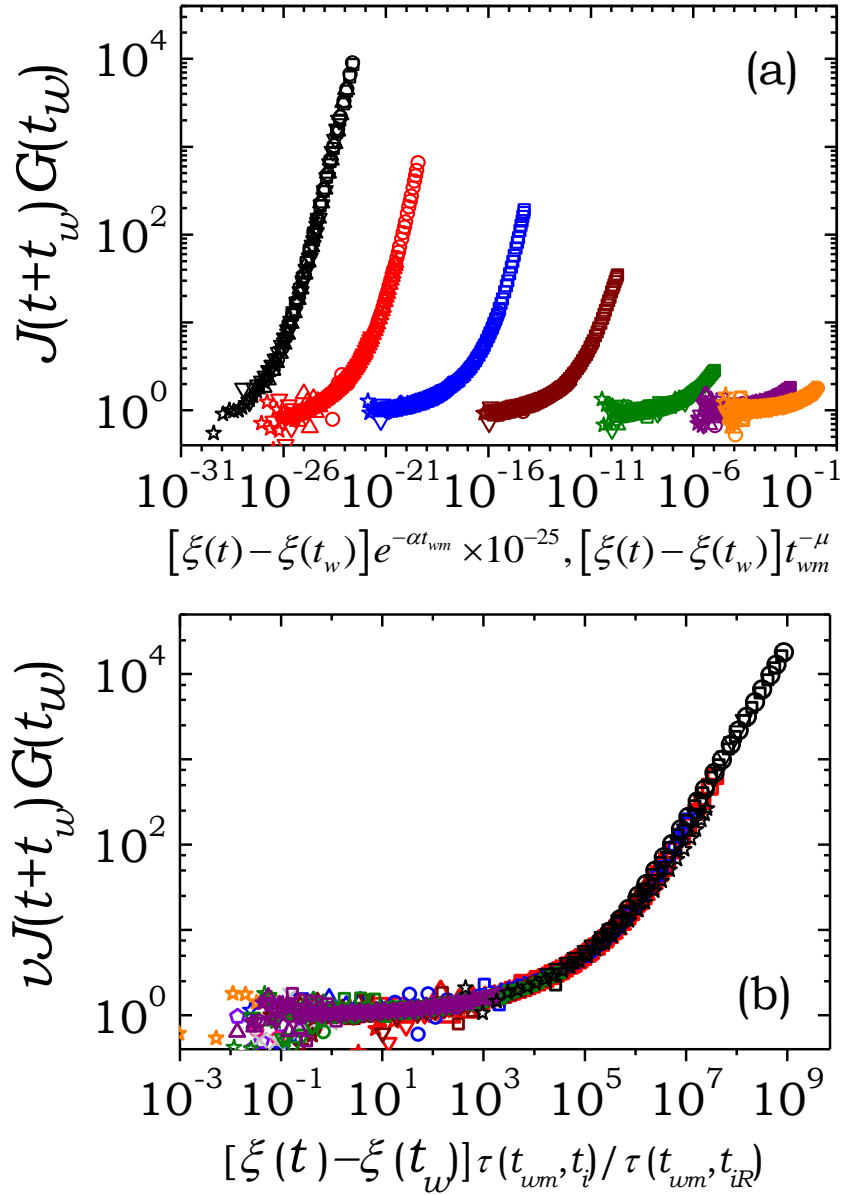

**Figure 3.** The time – aging time superpositions obtained using the effective time approach on various idle times is shown in figure (a) [from left to right: 13, 21, 30, 45, 56, 80, 180 day). Figure (b) represents comprehensive superposition, wherein all the creep curves are horizontally shifted on to a creep curve belonging to 45 days. Vertical shift factor $v$ in figure b is used to carry out vertical adjustment and is always in the range 1 – 1.2.

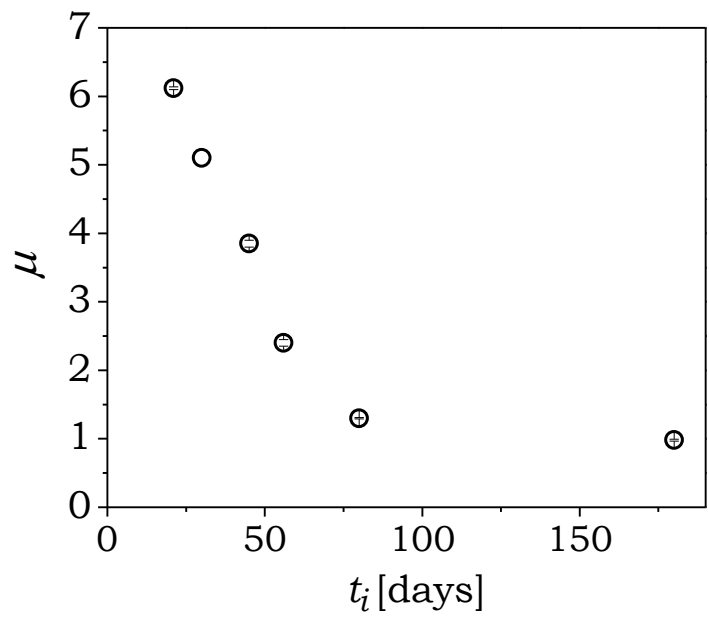

**Figure 4.** Parameter $\mu$ necessary to obtain superposition shown in figure 3 (a) plotted as a function of idle time.

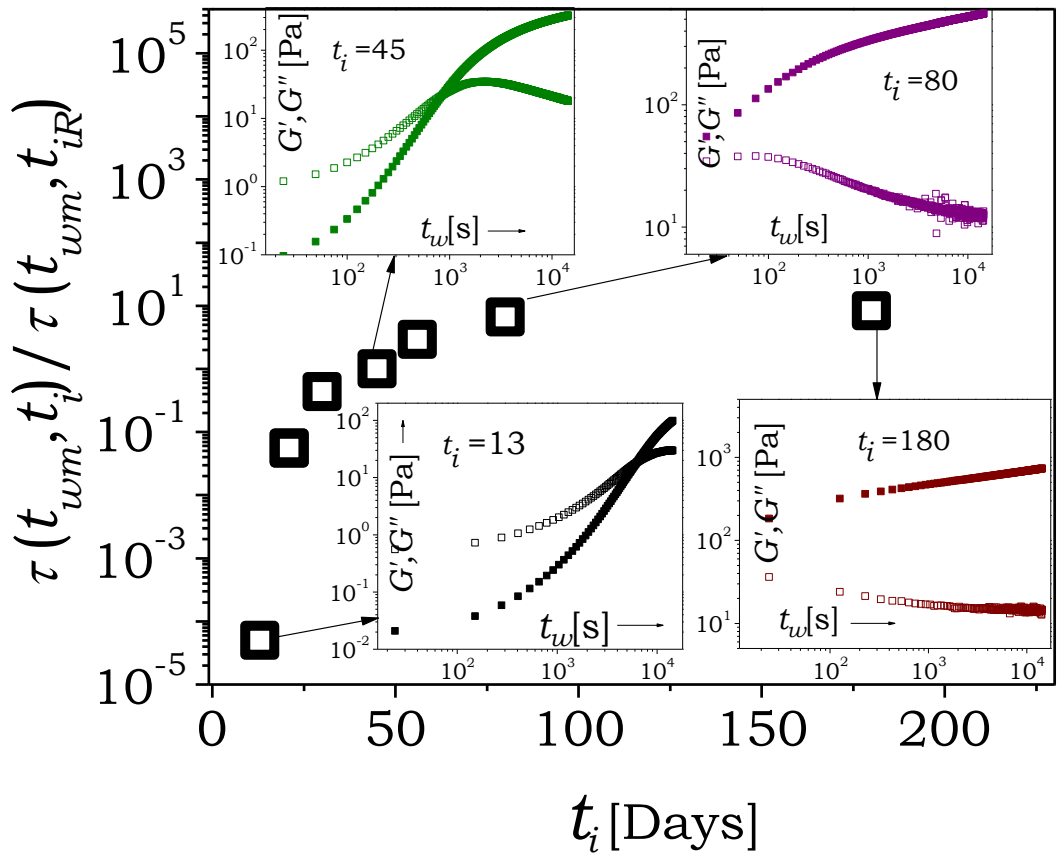

**Figure 5.** Ratio of relaxation time at $t_{wm}$ =14400 s on respective idle times to that at reference idle time plotted as a function of idle time. The various insets represent evolution of elastic and viscous modulus at respective idle times.

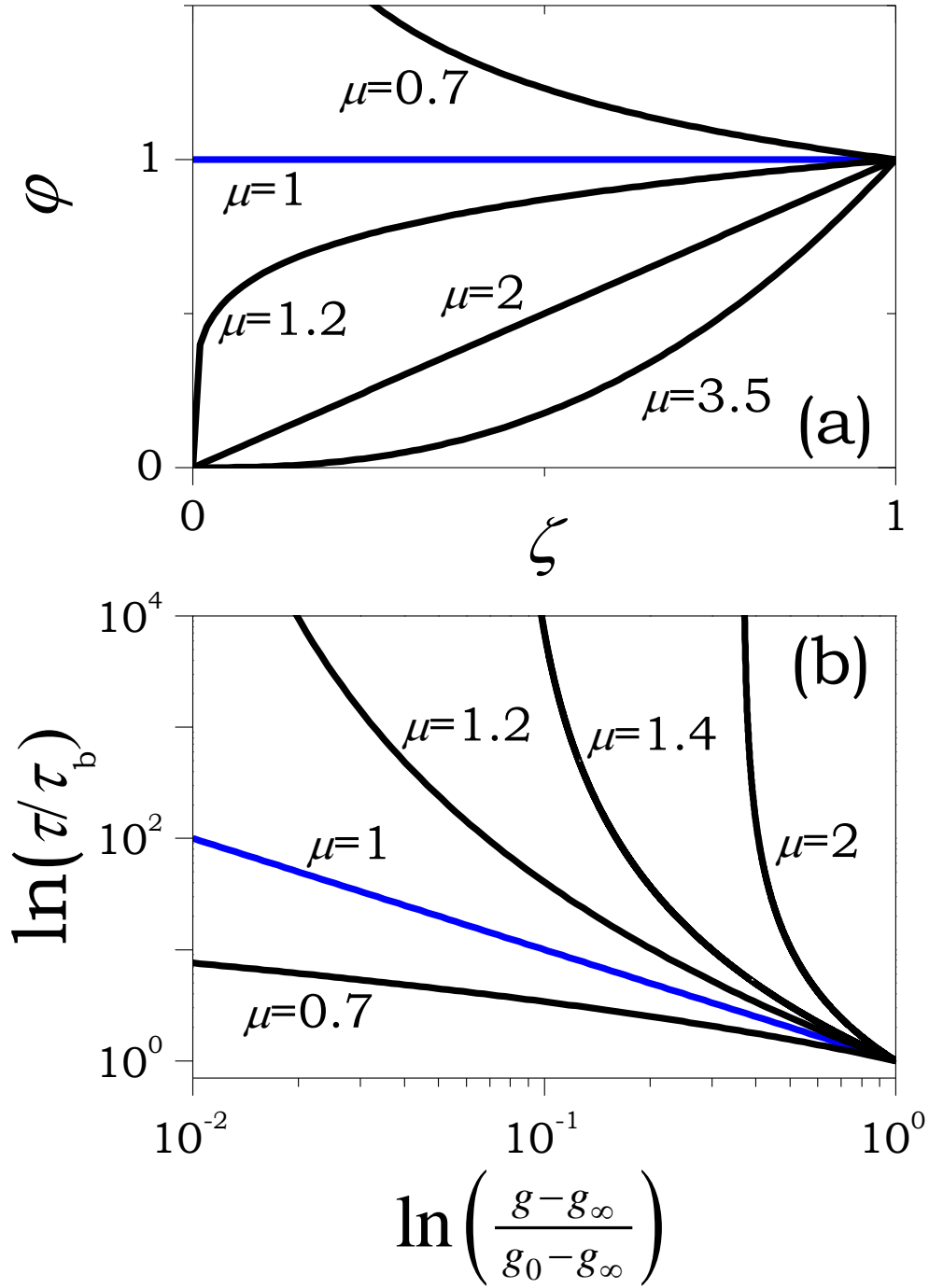

**Figure 6.** A schematic representing (a) dependence of $\varphi$ on dimensionless time $\zeta$ [equation (14)] and (b) dependence of relaxation time on excess Gibbs free energy [equation (15) with $A=1$ and $\tau_m = \tau_b$].

# Supplementary information

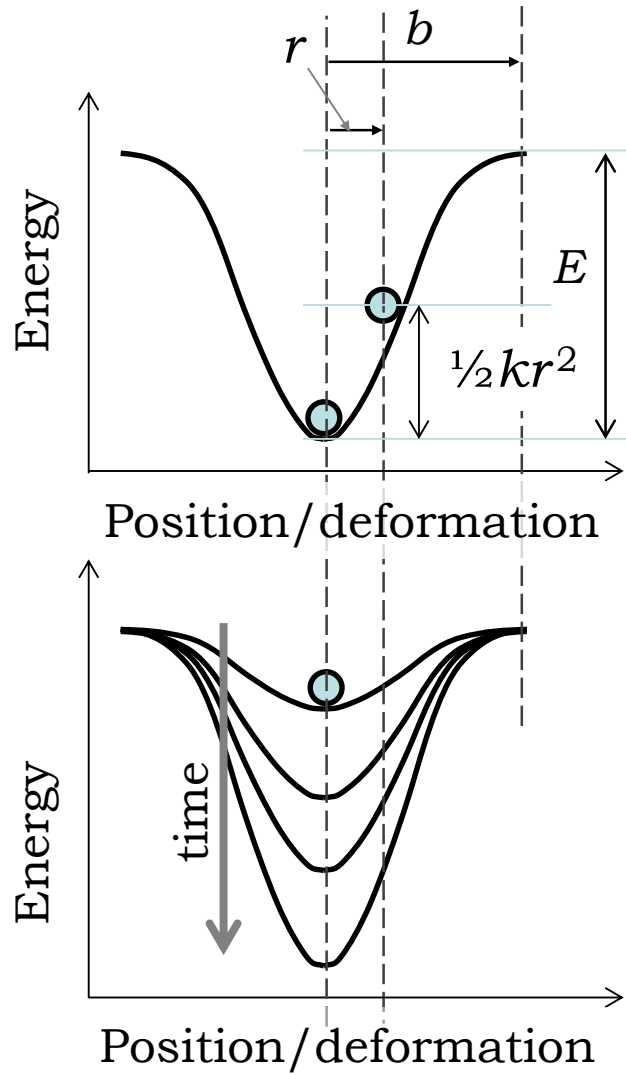

**Figure S1.** Schematic representing particle trapped in energy well under a deformation field. Bottom schematic describes aging of a well in which well depth goes on increasing as a function of time.

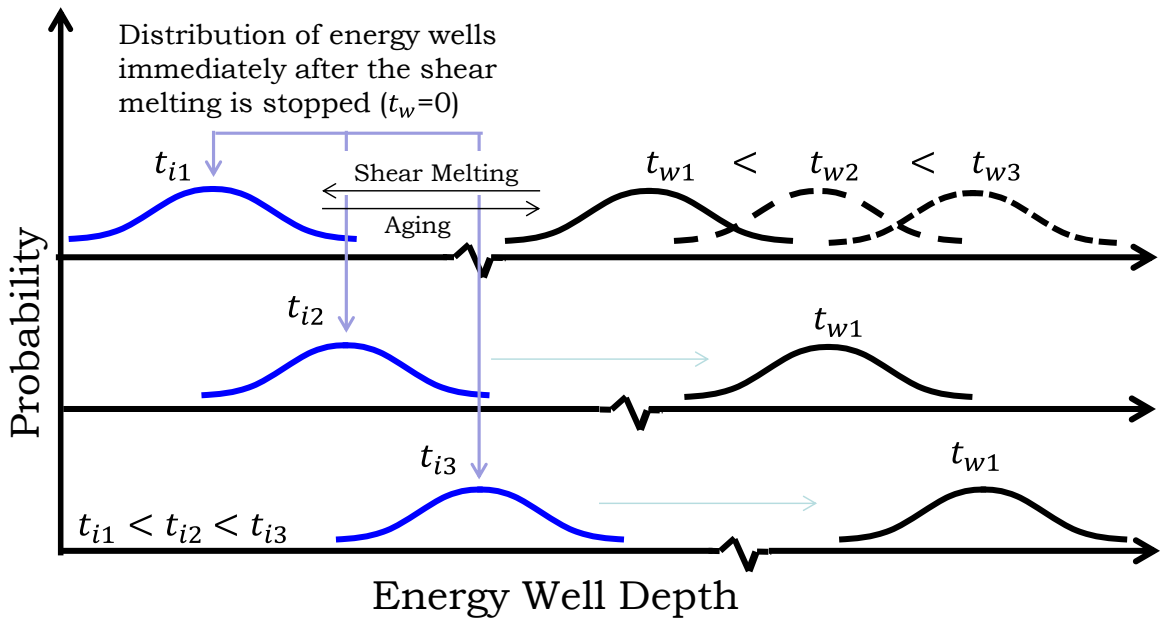

**Figure S2.** Effect of shear melting at different idle times on energy well distribution. Owing to irreversible aging, shear melting on greater idle times shifts the energy well distribution dome to higher well depths. Consequently suspension has greater average relaxation time for experiments carried out at same aging times but greater idle times. Self-similarity of the shape of the distribution dome as a function of aging time and idle time is a necessary condition to observe superpositions shown in figure 3.